\documentclass{article}
\usepackage{graphicx}
\usepackage{latexsym}

\def\beq{\begin{equation}}
\def\eeq{\end{equation}}
\def\rmd{{\rm d}}

\begin{document}

\title{\bf \LARGE Periastron shift in Weyl class spacetimes
}

\author{
Donato Bini$^* {}^\S{}^\P$,
Francesco De Paolis$^\dagger$,
Andrea Geralico$^\dagger {}^\S$, \\
Gabriele Ingrosso$^\dagger$
and Achille Nucita$^\dagger$
\\[4mm]%
{\small \it $^*$ Istituto per le Applicazioni del Calcolo ``M. Picone'', CNR I-00161 Rome, Italy}\\
{\small \it $\S$ International Center for Relativistic Astrophysics - I.C.R.A.,}\\
{\small \it University of Rome ``La Sapienza'', I-00185 Rome, Italy}\\%
{\small \it $\P$ INFN - Sezione di Firenze, Polo Scientifico, via Sansone, 1}\\
{\small \it  I-50019, Sesto Fiorentino (FI), Italy }\\
{\small \it $^\dagger$ Dipartimento di Fisica, Universit\`a di Lecce, and INFN - Sezione di Lecce,} \\
{\small \it Via Arnesano, CP 193, I-73100 Lecce, Italy}\\
}

\date{{\small \today }}

\maketitle

\begin{abstract}
The periastron position advance for geodesic motion in axially
symmetric solutions of the Einstein field equations belonging to
the Weyl class of vacuum solutions is investigated. Explicit
examples corresponding to either static solutions (single
Chazy-Curzon, Schwarzschild and a pair of them), or stationary
solution (single rotating Chazy-Curzon and Kerr black hole) are
discussed. The results are then applied to the case of S2-SgrA$^*$
binary system of which the periastron position advance will be
soon measured with a great accuracy.

\vspace*{5mm}
\noindent  \\
Keywords: Periastron shift, Weyl class spacetimes.\\
PACS number: 04.20.Cv
\end{abstract}
\thispagestyle{empty}

\section{Introduction}

Stellar sources close enough to a massive central body will likely
not be on simple Keplerian orbit due to general relativistic
effects. In particular, a periastron position shift will result in
(observable) {\textit {rosetta}} shaped orbits. Since the amount
of periastron advance strongly depends on the compactness of the
central body, the detection of such an effect will give
information about the nature of the central body itself. This
could be exactly the case of stars orbiting close to the center of
our Galaxy, where a \lq\lq dark object" is  presumably hosted. All the
literature concerning this topic considers the central body as a
static Schwarzschild black hole and a wide variety of related
results as well as estimates for measurable shift effects can be
found.

However, General Relativity contains other interesting exact
solutions representing naked singularities or superposition of two
or more axially symmetric bodies kept apart on stable
configuration by gravitationally inert singular structures whose
observational aspect have been poorly examined or often dismissed as representative of non-physical situations.
On the other hand, singularities are somehow typical in general relativity and it is especially in order to better understand their role and their proper character in this quite simple class of solutions that  the present paper has been conceived. 
In particular,
together with the well known Schwarzschild black hole solution one
may consider the Chazy-Curzon solution which represents the static
exterior gravitational field of a deformed mass endowed with a
naked singularity at the particle position \cite{chazy,curzon}.
Analogously, for the rotating case, together with the well known
Kerr black hole solution  one may consider the rotating
Chazy-Curzon case.

Furthermore, due to a proper use of the Weyl formalism, it is easy
to extend the discussion including  the superposition of two or
more axially symmetric bodies (in particular, two non-rotating
Chazy-Curzon particles or Schwarzschild black holes) aligned along
the $z-$axis in a static configuration. These solutions are
characterized by the occurrence of a conical singularity on the
$z-$axis, which can be thought as a \lq\lq strut", keeping the two
bodies apart at a fixed distance (see also the pioneering works of 
Semer\'ak, Zellerin and ${\check {\rm Z}}$\'a${\check {\rm c}}$ek \cite{sem1,sem2}
which considered accretion disks or rings of astrophysical interest).

The aim of the paper is to estimate the periastron shift of a star
orbiting one of the above mentioned object and compare the results
obtained. In doing so, we consider the S2-SgrA$^*$ binary
system, hosted in the center of our Galaxy, which is of special
interest in view of future observations possibly revealing a
periastron shift effect.

The paper is structured as follows: in Section 2 we introduce some
static metric in Weyl cylindrical coordinates: the single
Chazy-Curzon particle, the single Schwarzschild black hole, two
Chazy-Curzon particles and two Schwarzschild black holes,
representing the gravitational field of a central body (a single
mass located at the origin of the coordinates, or two masses
displaced along the $z-$axis). In Section 3 we study the geodesic
equations in such spacetimes in order to obtain a (linearized)
expression for the periastron  advance, up to the second order in
a mass parameter associated to the central object (which exactly
coincides with the mass if the central object is a single body; it
is, instead, related to both the masses when the central object
consists in a superposition of two bodies). In Section 4, we
extend our treatment to the stationary case of a Kerr black hole
as well as a rotating Chazy-Curzon particle. We apply our results
to the case of S2-SgrA$^*$ binary system in Section 5, and in
Section 6 we address some conclusions.

\section{Axisymmetric, static, vacuum solutions in Weyl formalism}

Axisymmetric, static, vacuum solutions of the Einstein field
equations can be described by the Weyl formalism \cite{weyl}. The
line element in cylindrical coordinates $(t,\rho,z,\phi)$ writes
as
\begin{equation}
\label{weylmetric}
\rmd s^2=-e^{2\psi}\rmd t^2+e^{2(\gamma-\psi)}[\rmd\rho^2+\rmd z^2]+\rho^2e^{-2\psi}\rmd\phi^2\
,
\end{equation}
where the functions $\psi$ and $\gamma$ depend on the coordinates
$\rho$ and $z$ only. The vacuum Einstein field equations in Weyl
coordinates reduce to
\begin{eqnarray}
\label{einsteqs}
0&=&\psi_{,\rho\rho}+\frac1\rho\psi_{,\rho}+\psi_{,zz}\ ,\nonumber\\
0&=&\gamma_{,\rho}-\rho[\psi_{,\rho}^2-\psi_{,z}^2]\ ,\nonumber\\
0&=&\gamma_{,z}-2\rho\psi_{,\rho}\psi_{,z}\ .
\end{eqnarray}
The first equation is the three-dimensional Laplace equation
written in cylindrical coordinates; so the function $\psi$ plays
the role of a newtonian potential. The linearity of that equation
allows to find explicit solutions representing superpositions of
two or more axially symmetric bodies. In general, these solutions
correspond to configurations not gravitationally stable because of
the occurrence of gravitationally inert singular structures
(``struts'' and ``membranes'') that keep the bodies apart (see,
e.g. \cite{letelier} and references therein). Properly speaking,
this is the effect of the presence of a conical singularity on the
$z-$axis, the occurrence of which is related to the non-vanishing
of the function $\gamma(\rho,z)$ on the portion of the axis
between the sources or outside them.

For the static axisymmetric vacuum solutions the regularity condition on the axis of symmetry (``elementary flatness'') is given by
\begin{equation}
\label{regcond}
\lim_{\rho\rightarrow0}\gamma=0\ .
\end{equation}

Let us briefly summarize the properties of the solutions belonging
to the Weyl class representing the field of a Chazy-Curzon
particle or a Schwarzschild black hole as well as superpositions
of two of them.

\begin{enumerate}

\item[${\mathit 1}$.] {\textit {The single Chazy-Curzon particle}}

A single Chazy-Curzon particle is a static axisymmetric solution of Einstein's equations endowed with a naked singularity at the particle position \cite{chazy,curzon,scott}.
The Curzon metric is generated by the newtonian potential of a spherically symmetric point mass using the Weyl formalism:
\begin{equation}
\label{ccsolw}
\psi_{\rm C}=-\frac{M_{\rm C}}{R_{\rm C}}\ , \qquad \gamma_{\rm C}=-\frac12\frac{M_{\rm C}^2\rho^2}{R_{\rm C}^4}\ , \qquad R_{\rm C}=\sqrt{\rho^2+z^2}\ .
\end{equation}

\item[${\mathit 2}$.] {\textit {Superposition of two Chazy-Curzon particles}}

The solution corresponding to the superposition of two Chazy-Curzon particles with masses $M_{\rm C}$ and $m_{\rm C_b}$ and positions $z=0$ and $z=b$ on the $z$-axis respectively is given by metric (\ref{weylmetric}) with functions
\begin{equation}
\label{psigammaCCb}
\psi=\psi_{\rm C}+\psi_{\rm C_b}\ , \qquad \gamma=\gamma_{\rm C}+\gamma_{\rm C_b}+\gamma_{\rm CC_b}\ ,
\end{equation}
where $\psi_{\rm C}$, $\gamma_{\rm C}$ are defined by Eq.~(\ref{ccsolw}), while
\begin{equation}
\label{ccbsolw}
\psi_{\rm C_b}=-\frac{m_{\rm C_b}}{R_{\rm C_b}}\ , \qquad \gamma_{\rm C_b}=-\frac12\frac{m_{\rm C_b}^2\rho^2}{R_{\rm C_b}^4}\ , \qquad R_{\rm C_b}=\sqrt{\rho^2+(z-b)^2}\
\end{equation}
and $\gamma_{\rm CC_b}$ can be obtained by solving Einstein's equations (\ref{einsteqs}):
\begin{equation}
\gamma_{\rm CC_b}=2\frac{m_{\rm C_b}M_{\rm C}}{b^2}\frac{\rho^2+z(z-b)}{R_{\rm C_b}R_{\rm C}}+C\ .
\end{equation}
The value of the arbitrary constant $C$ can be determined by
imposing the regularity condition (\ref{regcond}). Note that in
order to make the function $\gamma_{\rm CC_b}$ vanishes on the
whole $z$-axis, the constant value $C$ cannot be uniquely chosen.
In fact, setting $\gamma_{\rm CC_b}\not= 0$ gives rise to a
conical singularity (see, e.g. \cite{sokolov,israel}),
corresponding to a strut in compression, which holds the two
particles apart. On the other hand, the choice $C=2m_{\rm
C_b}M_{\rm C}/b^2$ makes $\gamma_{\rm CC_b}=0$ only on the segment
$0<z<b$ of the $z$-axis between the sources. In the following we
use $C=-2m_{\rm C_b}M_{\rm C}/b^2$, that makes $\gamma_{\rm
CC_b}=0$ on the portion of the axis with $z<0$ and $z>b$.

\item[${\mathit 3}$.] {\textit {The single Schwarzschild black hole}}

The Schwarzschild black hole solution in Weyl coordinates is
generated by the newtonian potential of a line source (a
homogeneous rod) of mass $M_{\rm S}$ and lenght $2L$ (with the
further position $L=M_{\rm S}$), lying on the axis and placed
about the origin:
\begin{eqnarray}
\label{Ssolw}
\psi_{\rm S}&=&\frac12\ln{\left[\frac{R_1^{+}+R_1^{-}-2M_{\rm S}}{R_1^{+}+R_1^{-}+2M_{\rm S}}\right]}\ , \qquad
\gamma_{\rm S}=\frac12\ln{\left[\frac{(R_1^{+}+R_1^{-})^2-4M_{\rm S}^2}{4R_1^{+}R_1^{-}}\right]}\ , \nonumber\\
R_1^{\pm}&=&\sqrt{\rho^2+(z\pm M_{\rm S})^2}\ .
\end{eqnarray}
The usual line element in standard Boyer-Lindquist coordinates  $(t,r,\theta,\phi)$ is recovered by making the following transformation:
\begin{equation}
\label{daweylabl}
\rho=\sqrt{r^2-2M_{\rm S}r}\sin\theta\ , \qquad z=(r-M_{\rm S})\cos\theta\ .
\end{equation}

\item[${\mathit 4}$.] {\textit {Superposition of two Schwarzschild black holes}}

The solution corresponding to a linear superposition of two
Schwarzschild black holes with masses $M_{\rm S}$ and $m_{\rm
S_b}$ and positions $z=0$ and $z=b$ on the $z$-axis respectively
is given by metric (\ref{weylmetric}) with functions
\begin{eqnarray}
\label{psigammaSSb}
\psi=\psi_{\rm S}+\psi_{\rm S_b}\ , \qquad \gamma=\gamma_{\rm S}+\gamma_{\rm S_b}+\gamma_{\rm SS_b}\ ,
\end{eqnarray}
where $\psi_{\rm S}$, $\gamma_{\rm S}$ are defined by Eq.~(\ref{Ssolw}), while
\begin{eqnarray}
\label{SSbsol}
\psi_{\rm S_b}&=&\frac12\ln{\left[\frac{R_2^{+}+R_2^{-}-2m_{\rm S_b}}{R_2^{+}+R_2^{-}+2m_{\rm S_b}}\right]}\ , \qquad
\gamma_{\rm S_b}=\frac12\ln{\left[\frac{(R_2^{+}+R_2^{-})^2-4m_{\rm S_b}^2}{4R_2^{+}R_2^{-}}\right]}\nonumber\\
\gamma_{\rm SS_b}&=&\frac12\ln{\left[\frac{E_{(1^{+},2^{-})}E_{(1^{-},2^{+})}}{E_{(1^{+},2^{+})}E_{(1^{-},2^{-})}}\right]}+C \ , \qquad
E_{(1^{\pm},2^{\pm})}=\rho^2+R_1^{\pm}R_2^{\pm}+Z_1^{\pm}Z_2^{\pm}\nonumber\\
R_1^{\pm}&=&\sqrt{\rho^2+(Z_1^{\pm})^2}\ , \qquad
R_2^{\pm}=\sqrt{\rho^2+(Z_2^{\pm})^2}\nonumber\\
Z_1^{\pm}&=&z\pm M_{\rm S}\ , \qquad
Z_2^{\pm}=z-(b\mp m_{\rm S_b})\ ,
\end{eqnarray}
the function $\gamma_{\rm SS_b}$ being obtained by solving Einstein's equations (\ref{einsteqs}).
The value of arbitrary constant $C$ can be determined by imposing the regularity condition (\ref{regcond}); we make the choice $C=0$, so that the function $\gamma_{\rm SS_b}$ vanishes on the portions of the $z$-axis outside the sources (that is, for $z>b+m_{\rm S_b}$ and $z<-M_{\rm S}$).

\end{enumerate}

\section{Periastron shift of a distant orbiting star}

The geodesic motion of a test particle in a plane orthogonal to
the $z-$axis (i.e. $z=const$), from the metric (\ref{weylmetric})
is described by the equations
\begin{eqnarray}
\label{geodesics}
0 &=& \ddot{\phi} +\frac{2}{\rho}(1-\rho \psi_{,\rho})\dot{\rho}\dot{\phi}\, \nonumber \\
0 &=& e^{-2(\gamma-2\psi)}[\dot{t}^2 +\rho ^2 e^{-4\psi}\dot{\phi} ^2]\psi _{,z} -[\gamma_{,z}-\psi_{,z}] \dot{\rho} ^2\, \nonumber \\
0 &=& e^{-2(\gamma-2\psi)}[\psi _{,\rho} \dot{t}^2 -\rho e^{-4\psi} (1-\rho \psi _{,\rho})] \dot{\phi} ^2 +[\gamma_{,\rho}-\psi_{,\rho}]\dot{\rho}^2+\ddot{\rho}\, \nonumber \\
0 &=& \ddot{t} +2\psi_{,\rho}\dot{t}\dot{\rho}~,
\end{eqnarray}
where the dot represents differentiation with respect to the orbit proper time
parameter and the metric functions $\psi$ and $\gamma$ now depend only
on $\rho$.
When the $z=const$ plane is a symmetry plane (the $z=0$ plane for a single-body solution, or the middle plane in the case of two bodies of equal mass), this system is further simplified, since in this case $\psi_{,z}=0$, and so $\gamma_{,z}=0$ too.

Due to the Killing symmetries of the metric the conserved energy per unit mass ($E$) and angular momentum per unit mass ($L$) of the particle can be introduced:
\beq
E=e^{2\psi} \dot{t}, \qquad L=\rho^2e^{-2\psi}\dot{\phi}.
\eeq
Moreover, by using the timelike condition for the geodesic,
\begin{equation}
-e^{-2\psi}
E^2+e^{2(\gamma-\psi)}\dot{\rho}^2+\frac{e^{2\psi}}{\rho^2}L^2=-1~,
\end{equation}
the shape of the orbit (i.e. $\rho$ as
a function of $\phi$) is described  by
\begin{equation}
\frac{\rmd\rho}{\rmd\phi} =\rho^2
e^{-(\gamma+2\psi)}\left[\frac{1}{B^2}-\frac{e^{4\psi}}{\rho ^2} -
\frac{e^{2\psi}}{L^2}\right]^{1/2}~, \label{drhodphi}
\end{equation}
being $B=L/E$.

Let us assume the test particle bound in an elliptic orbit
around the compact object (which may be a
single Chazy-Curzon particle, a single Schwarzschild black hole or
a pair of them) on a plane orthogonal to the $z-$axis. The orbit followed
by the test particle can be parametrized as
\begin{equation}
\rho = \frac{d(1-e^2)}{1+e\cos \chi} \label{equationofellipse}~,
\end{equation}
where $d$ and $e$ can be thought as the ellipse semi-major axis
and eccentricity, and where $\chi$ is a new variable called {\it
relativistic anomaly}. At periastron and aphastron, $\rho$ reaches
its minimum ($\rho_ - = d(1-e)$) and maximum ($\rho_ + = d(1+e)$)
values obtained from eq. (\ref{equationofellipse}) for $\chi =0$
and $\chi=\pi$ respectively. At these points $\rmd\rho/\rmd\phi$
vanishes so that eq. (\ref{drhodphi}) gives
\begin{equation}
\left[\frac{e^{4\psi _{\pm}}}{\rho_{\pm} ^2} -
\frac{e^{2\psi_{\pm}}}{L^2}\right] = \frac{1}{B^2}~,
\end{equation}
with $\psi_{\pm} = \psi(\rho_{\pm})$. From the two previous
equations one can derive the two constant of motion to be
\begin{equation}
B^2 = \frac{e^{-2\psi_-} -
e^{-2\psi_+}}{\displaystyle{\left[\frac{e^{2\psi_-}}{\rho_ - ^2} -
\frac{e^{2\psi_+}}{\rho_ + ^2}\right]}}~, \qquad L^2 =\left[
\frac{e^{-2\psi_-}}{B^2} - \frac{e^{2\psi_-}}{{\rho_ -
^2}}\right]^{-1}~.
\end{equation}
It is  useful  to
express $\phi$ as a function of the relativistic anomaly $\chi$,
so that
\begin{equation}
\frac{\rmd\phi}{\rmd\chi}=\frac{d(1-e^2)e\sin \chi}{(1+e\cos
\chi)^2}\frac{\rmd\phi}{\rmd\rho}\bigg\vert_{\rho=\rho(\chi)}~,
\end{equation}
where $\rmd\phi/\rmd\rho$ can be derived from eq. (\ref{drhodphi})
by using the relation $\rho=\rho(\chi)$ of
eq. (\ref{equationofellipse}). The total change in $\phi$ as $\chi$ decreases from $\pi$ to 0 is
the same as the change in $\phi$ as $\chi$ increases from 0 to
$\pi$, so that the total change in $\phi$ per revolution is
$2|\phi(\pi)-\phi(0)|$, where
\begin{equation}
\label{integralephi}
\phi(\pi)-\phi(0)=\int_0^{\pi}\frac{d(1-e^2)e\sin \chi}{(1+e\cos
\chi)^2}\frac{\rmd\phi}{\rmd\rho}\bigg\vert_{\rho=\rho(\chi)}~\rmd\chi~.
\end{equation}
This would equal $2\pi$ if the orbit is a
closed ellipse, so in general the orbit precesses
by an angle
\begin{equation}
\label{precession}
\Delta \phi = 2|\phi(\pi)-\phi(0)| -2\pi~
\end{equation}
per revolution.
The integral appearing at the second hand in eq.
(\ref{integralephi}) can be generally expressed in terms of elliptic
integrals. However, by noting that the ratio between the total mass $M$
of the central object and the
major axis $d$ of the considered orbit is a very small quantity,
it is justifiable to expand the integral in powers of $M$,
retaining only the first power.

We are interested here in the estimate of the differences in the periastron position advance
due to the presence of a central massive object consisting in a single or two
Chazy-Curzon particles or Schwarzschild black holes, in order to priviledge either configuration
when a physical system (for instance the S2-SgrA$^*$ binary system) is investigated.
As we shall see soon, all the considered solutions give rise to the same amount of periastron shift at first
order in $M$, while differences occurr in the second order term.
In the case of a single Chazy-Curzon particle and a Schwarzschild black hole, by using the metric
functions (\ref{ccsolw}) and (\ref{Ssolw}) respectively evaluated at $z=0$, the integral (\ref{integralephi}) gives
\begin{eqnarray}
\label{1cc1schshift}
\Delta \phi_{\rm C} &\simeq& \frac{6\pi M_{\rm C}}{d(1-e^2)}+\frac{(44-9e^2)\pi M_{\rm C}^2}{2d^2(1-e^2)^2}\ , \nonumber\\
\Delta \phi_{\rm S}&\simeq& \frac{6\pi M_{\rm S}}{d(1-e^2)}+\frac{3(14-3e^2)\pi M_{\rm S}^2}{2d^2(1-e^2)^2}~.
\end{eqnarray}
Therefore, the two cases differ for the second order term; if we
take $M_{\rm C}=M=M_{\rm S}$, the difference reads out to be 
\beq
\label{1ccmen1schw}
\Delta \phi_{\rm C} - \Delta \phi_{\rm
S}= \frac{\pi M^2}{d^2(1-e^2)^2}\equiv \Delta\ . 
\eeq 
As far as a system of two
axially symmetric bodies is concerned, we limit our analysis to
the symmetric configuration consisting of two equal object endowed
each with the same mass $M$ displaced along the $z-$axis and
separated by a distance $b$. It has been demonstrated
\cite{biniger} that in this special case timelike circular
geodesics exist on the middle plane $z=b/2$. So we apply the above
developed formalism to a test particle which orbits a system of
two equal mass Chazy-Curzon singularities ($m_{\rm C_b}=M_{\rm
C}$) or Schwarzschild black holes ($m_{\rm S_b}=M_{\rm S}$) on the
middle plane. In both cases we solve the integral
(\ref{integralephi}) by expanding the argument also in the
distance parameter $b$ up to the second order. The expected
periastron shift results to be
\begin{eqnarray}
\label{2cc2schshift}
\Delta \phi_{\rm CC_b} &\simeq& \frac{12\pi M_{\rm C}}{d(1-e^2)}+\frac{2(44-9e^2)\pi M_{\rm C}^2}{d^2(1-e^2)^2}-\frac{3}{4}\frac{\pi b^2}{d^2(1-e^2)^2}~, \nonumber \\
\Delta \phi_{\rm SS_b} &\simeq& \frac{12\pi M_{\rm S}}{d(1-e^2)}+\frac{3(29-6e^2)\pi M_{\rm S}^2}{d^2(1-e^2)^2}-\frac{3}{4}\frac{\pi b^2}{d^2(1-e^2)^2}~.
\end{eqnarray}
Note that the solutions for a single Chazy-Curzon particle or
Schwarzschild black hole in $z=0$ can be recovered from the
previous relations by putting $b\rightarrow 0$ and $M\rightarrow
M/2$ in the first case and $b\rightarrow 2M$ and then
$M\rightarrow M/2$ in the latter case, respectively. The
difference in the periastron shift exactly coincides with that
estimated for the corresponding single-body solutions:
\beq
\label{2ccmen2schw} 
\Delta \phi_{\rm CC_b} - \Delta
\phi_{\rm SS_b}= \Delta\ , 
\eeq 
by taking $M_{\rm C}=M=M_{\rm S}$, with $\Delta$ defined by (\ref{1ccmen1schw}).

It is interesting to point out that the second order term appearing in both formulae (\ref{2cc2schshift}) can be made vanishing by a suitable choice of the distance $b$
between the two bodies: by introducing the distance parameter $\beta=b/M$, we find
\beq
\beta_{\rm CC_b}=\frac23\sqrt{6}\sqrt{44-9e^2}\ , \qquad \beta_{\rm SS_b}=2\sqrt{29-6e^2}\ .
\eeq

\section{Axisymmetric, stationary, vacuum solutions in Weyl formalism}

The line element of a stationary axisymmetric spacetime is given
by the Lewis-Papapetrou metric \cite{exactsols}
\begin{equation}
\label{lewispapmetric}
\rmd s^2=-e^{2\psi}[\rmd t-w\rmd\phi]^2+e^{2(\gamma-\psi)}[\rmd\rho^2+\rmd z^2]+\rho^2e^{-2\psi}\rmd\phi^2\
,
\end{equation}
where the function $\psi$, $\gamma$ and $w$ depend on the coordinates
$\rho$ and $z$ only. The vacuum Einstein field equations in Weyl
coordinates reduce to
\begin{eqnarray}
\label{einsteqskerr}
0&=&\psi_{,\rho\rho}+\frac1\rho\psi_{,\rho}+\psi_{,zz}+\frac{e^{4\psi}}{2\rho^2}[w_{,\rho}^2+w_{,z}^2]\ , \nonumber\\
0&=&w_{,\rho\rho}-\frac1\rho w_{,\rho}+w_{,zz}+4[w_{,\rho}\psi_{,\rho}+w_{,z}\psi_{,z}]\ , \nonumber\\
0&=&\gamma_{,\rho}-\rho[\psi_{,\rho}^2-\psi_{,z}^2]+\frac{e^{4\psi}}{4\rho}[w_{,\rho}^2-w_{,z}^2]\ , \nonumber\\
0&=&\gamma_{,z}-2\rho\psi_{,\rho}\psi_{,z}+\frac{e^{4\psi}}{2\rho}w_{,\rho}w_{,z}\ .
\end{eqnarray}
The metric function $\gamma$ is obtained by quadrature once the
solutions $\psi$ and $w$ for the nonlinear coupled system of the
first two equations above are known.
The geodesic equations for the metric (\ref{lewispapmetric}) (when the motion is confined on a $z=const$ plane) are given by
\begin{eqnarray}
0 &=& \ddot{t}+\left[2\psi _{,\rho} +\frac{e^{4\psi}}{\rho ^2} w w_{,\rho}\right]\dot{t}\dot{\rho} +\left[-4\psi _{,\rho}w-\left(1+\frac{e^{4\psi}}{\rho^2}w^2\right)w_{,\rho}+\frac{2}{\rho}w\right]\dot{\rho}\dot{\phi}\ , \nonumber \\
0 &=& e^{-2(\gamma -2\psi)}\{\psi_{,z} \dot{t} ^2 - (w_{,z} +2w\psi_{,z})\dot{\phi}\dot{t}+ [w(w_{,z}+w\psi_{,z})+e^{-4\psi}\rho ^2 \psi_{,z}]\dot{\phi} ^2\} \nonumber \\
&&+ [\psi_{,z}-\gamma _{,z}]\dot{\rho} ^2\ ,\nonumber \\
0 &=& e^{-2(\gamma -2\psi)}\{\psi_{,\rho} \dot{t} ^2 - (w_{,\rho}+2w\psi_{,\rho})\dot{\phi}\dot{t}+ [w(w_{,\rho}+w\psi_{,\rho})-e^{-4\psi}\rho (1-\rho \psi_{,\rho})]\dot{\phi} ^2\}\nonumber \\
&&+[\gamma_{,\rho}-\psi _{,\rho}]\dot{\rho}^2+\ddot{\rho}\ , \nonumber \\
0 &=& \ddot{\phi}+\frac{e^{4\psi}}{\rho^2}w_{,\rho}\dot{t}\dot{\rho}+\left[\frac{2}{\rho}-\frac{e^{4\psi}}{\rho ^2}w w_{,\rho} -2\psi_{,\rho} \right]\dot{\rho}\dot{\phi}~.
\end{eqnarray}
Following the same procedure of the static case, from the previous set of equations we finally obtain
\begin{equation}
\frac{\rmd\rho}{\rmd\phi} =
\frac{e^{-(\gamma-2\psi)}}{1-\frac{2w}{B}}[e^{-4\psi}\rho ^2 -w
(e^{-2\psi} -4w)]
\left[\frac{1}{B^2}-\frac{\left(1-\frac{2w}{B}\right)^2}{e^{-4\psi}\rho
^2 -w (e^{-2\psi} -4w)} - \frac{e^{2\psi}}{L^2}\right]^{1/2},
\label{drhodphikerr}
\end{equation}
where the metric functions $\psi$, $\gamma$ and $w$ now depend only on $\rho$.
The periastron shift is given again by eq. (\ref{precession}), by substituting the previous expression for $\rmd\rho/\rmd\phi$ into equation (\ref{integralephi}).
As applications we shall consider next the solutions corresponding to the single Kerr black hole as well as the rotating Chazy-Curzon particle.
Superposition of two rotating solutions can also be considered. However, the involved formulae are rather complicated, and we shall not pursue this extension here.

\begin{enumerate}

\item[${\mathit 1}$.] {\textit {The Kerr black hole}}

The metric functions generating the Kerr black hole solution in Weyl coordinates are given by
\begin{eqnarray}
\label{Ksolw}
\psi_{\rm K}&=&\frac12\ln{\left[\frac{(R_{\rm K}^{+}+R_{\rm K}^{-})^2-4M_{\rm K}^2+\frac{a^2}{\sigma^2}(R_{\rm K}^{+}-R_{\rm K}^{-})^2}{[R_{\rm K}^{+}+R_{\rm K}^{-}+2M_{\rm K}]^2+\frac{a^2}{\sigma^2}(R_{\rm K}^{+}-R_{\rm K}^{-})^2}\right]}\ , \nonumber\\
\gamma_{\rm K}&=&\frac12\ln{\left[\frac{(R_{\rm K}^{+}+R_{\rm K}^{-})^2-4M_{\rm K}^2+\frac{a^2}{\sigma^2}(R_{\rm K}^{+}-R_{\rm K}^{-})^2}{4R_{\rm K}^{+}R_{\rm K}^{-}}\right]}\ , \nonumber\\
w_{\rm K}&=&-\frac{aM_{\rm K}}{\sigma^2}\frac{[R_{\rm K}^{+}+R_{\rm K}^{-}+2M_{\rm K}][(R_{\rm K}^{+}-R_{\rm K}^{-})^2-4\sigma^2]}{(R_{\rm K}^{+}+R_{\rm K}^{-})^2-4M_{\rm K}^2+\frac{a^2}{\sigma^2}(R_{\rm K}^{+}-R_{\rm K}^{-})^2}\ ,
\end{eqnarray}
where
\beq
R_{\rm K}^{\pm}=\sqrt{\rho^2+(z\pm\sigma)^2}\ , \qquad \sigma=\sqrt{M_{\rm K}^2-a^2}\ ,
\eeq
$M_{\rm K}$ and $a$ being the mass and the specific angular momentum of the source, respectively.
The usual line element in Boyer-Lindquist coordinates
$(t,r,\theta,\phi)$ is recovered by making the transformation
\begin{equation}
\rho=\sqrt{r^2-2M_{\rm K}r+a^2}\sin\theta\ , \qquad
z=(r-M_{\rm K})\cos\theta\ . \label{daweylakerr}
\end{equation}

Following the procedure described in the previous section, the
periastron shift of a star orbiting a Kerr black hole is given by
\begin{equation}
\label{1kshift}
\Delta \phi_{\rm K} \simeq  \frac{6\pi M_{\rm K}}{d(1-e^2)} +\frac{16a\pi M_{\rm K}^{1/2}}{d^{3/2}(1-e^2)^{3/2}}
+\frac{3(14-3e^2)\pi M_{\rm K}^2}{2d^2(1-e^2)^2} +\frac{3\pi a^2}{d^2(1-e^2)^2}~,
\end{equation}
where we have expanded the result of the integral also to the second order
in $a$. Obviously this result reduces to
the Schwarzschild one (see eq. (\ref{1cc1schshift})) for
$a=0$, as expected.

\item[${\mathit 2}$.] {\textit {The rotating Chazy-Curzon particle}}

A stationary generalization of the Curzon solution (\ref{ccsolw}) representing  a rotating Chazy-Curzon particle is given by the metric (\ref{lewispapmetric}) with functions (see \cite{manko} and references therein)
\begin{eqnarray}
\label{ccrotsolw}
\psi_{\rm C_{rot}}&=&\frac12\ln{\left[e^{2\psi_{\rm C}}\frac{F_1}{F_2}\right]}\ , \qquad \gamma_{\rm C_{rot}}=\frac12\ln{\left[\frac{K_1}{16}e^{2\gamma_{\rm C}}\frac{F_1}{R_{\rm C_{rot}}^{+\ 4}R_{\rm C_{rot}}^{-\ 4}}\right]}\ , \nonumber\\
w_{\rm C_{rot}}&=&-ke^{-2\psi_{\rm C}}\frac{F_3}{F_1}+K_2\ , \qquad R_{\rm C_{rot}}^{\pm}=\sqrt{\rho^2+(z\pm k)^2}\ ,
\end{eqnarray}
where $\psi_{\rm C}$, $\psi_{\rm C}$ refer to the seed solution (\ref{ccsolw}),  $F_1, F_2, F_3$ are functions of the coordinates $\rho$ and $z$
\begin{eqnarray}
F_1&=&\{16R_{\rm C_{rot}}^{+\ 2}R_{\rm C_{rot}}^{-\ 2}+[(R_{\rm C_{rot}}^{+}+R_{\rm C_{rot}}^{-})^2-4k^2]^2a_1a_2\}^2\nonumber\\
&&+[(R_{\rm C_{rot}}^{+}+R_{\rm C_{rot}}^{-})^2-4k^2][(R_{\rm C_{rot}}^{+}-R_{\rm C_{rot}}^{-})^2-4k^2](R_{\rm C_{rot}}^{+\ 2}a_1-R_{\rm C_{rot}}^{-\ 2}a_2)^2\ , \nonumber\\
F_2&=&\{16R_{\rm C_{rot}}^{+\ 2}R_{\rm C_{rot}}^{-\ 2}+[(R_{\rm C_{rot}}^{+}+R_{\rm C_{rot}}^{-})^2-4k^2][R_{\rm C_{rot}}^{+}+R_{\rm C_{rot}}^{-}+2k]^2a_1a_2\}^2\nonumber\\
&&+[R_{\rm C_{rot}}^{+}+R_{\rm C_{rot}}^{-}+2k]^2\{-R_{\rm C_{rot}}^{+\ 2}[R_{\rm C_{rot}}^{+}-R_{\rm C_{rot}}^{-}-2k]a_1\nonumber\\
&&+R_{\rm C_{rot}}^{-\ 2}[R_{\rm C_{rot}}^{+}-R_{\rm C_{rot}}^{-}+2k]a_2\}^2\ , \nonumber\\
F_3&=&-R_{\rm C_{rot}}^{+\ 3}R_{\rm C_{rot}}^{-\ 3}[R_{\rm C_{rot}}^{+}+R_{\rm C_{rot}}^{-}+2k]\{R_{\rm C_{rot}}^{+}[R_{\rm C_{rot}}^{+}-R_{\rm C_{rot}}^{-}-2k]a_1\nonumber\\
&&+R_{\rm C_{rot}}^{-}[R_{\rm C_{rot}}^{+}-R_{\rm C_{rot}}^{-}+2k]a_2+[(R_{\rm C_{rot}}^{+}+R_{\rm C_{rot}}^{-})^2-4k^2]\nonumber\\
&&\times[R_{\rm C_{rot}}^{+}+R_{\rm C_{rot}}^{-}+2k]^2a_1a_2\{[R_{\rm C_{rot}}^{+}-R_{\rm C_{rot}}^{-}-2k](a_1/R_{\rm C_{rot}}^{-\ 3})\nonumber\\
&&+[R_{\rm C_{rot}}^{+}-R_{\rm C_{rot}}^{-}+2k](a_2/R_{\rm C_{rot}}^{+\ 3})\}\}\ ,
\end{eqnarray}
with
\beq
a_1=\alpha\, e^{-2[R_{\rm C_{rot}}^{-}/R_{\rm C}-1]}\ , \qquad a_2=\alpha\, e^{-2[R_{\rm C_{rot}}^{+}/R_{\rm C}-1]}\ ,
\eeq
$R_{\rm C}$ is given in (\ref{ccsolw}), and $K_1, K_2, k, \alpha$ are real constants, which are fixed by requiring a regular behaviour of the metric functions $\gamma_{\rm C_{rot}}$ and $w_{\rm C_{rot}}$ on the symmetry axis,
in order to satisfy the condition of asymptotic flatness for the solution. They are expressed in terms of $k$ and $\alpha$, which define the total mass $M_{\rm C_{rot}}$
and the total angular momentum $J_{\rm C_{rot}}$ of the rotating source
\beq
\label{MJCrot}
M_{\rm C_{rot}}=k\frac{1-3\alpha^2}{1-\alpha^2}\ , \qquad J_{\rm C_{rot}}=2\alpha k^2\frac{3-5\alpha^2}{(1-\alpha^2)^2}\ ,
\eeq
as
\beq
\label{K12}
K_1=\frac1{(1-\alpha^2)^2}\ , \qquad K_2=\frac{4k\alpha}{1-\alpha^2}\ .
\eeq
Eqs. (\ref{ccrotsolw}) - (\ref{K12}) fully determine the metric which could be used for the description of the exterior field of a stationary Curzon mass.
When the rotation parameter $\alpha=0$, it reduces to the static Curzon solution (\ref{ccsolw}).
By requiring that the total mass $M_{\rm C_{rot}}$ be a positive quantity, one obtains the admissible values of the rotation parameter $\alpha$:
\beq
|\alpha|<\frac12\ , \qquad {\mbox {\rm and}} \qquad |\alpha|>1\ .
\eeq
In the limit of small values of the rotation parameter, the relation (\ref{MJCrot}) defining the total mass and angular momentum of the source becomes
\beq
M_{\rm C_{rot}}\simeq k\ , \qquad J_{\rm C_{rot}}\simeq 6\alpha M_{\rm C_{rot}}^2\ ,
\eeq
and the expected periastron shift results to be given by
\begin{equation}
\label{1ccrotshift}
\Delta \phi_{\rm C_{rot}} \simeq  \frac{6\pi M_{\rm C_{rot}}}{d(1-e^2)} -\frac{96\alpha\pi M_{\rm C_{rot}}^{3/2}}{d^{3/2}(1-e^2)^{3/2}}
+\frac{(44-9e^2)\pi M_{\rm C_{rot}}^2}{2d^2(1-e^2)^2}~,
\end{equation}
where terms beyond the second order in $1/d$ have been neglected, and reduces to the corresponding one for the static case (see eq. (\ref{1cc1schshift})) for $\alpha=0$.

\end{enumerate}

\section{Application: the S2-SgrA$^*$ binary system}

Recently Ghez et al. \cite{ghezetal} have observed a star orbiting
close to the galactic center massive black hole. The star, which
has been labelled as S2, with mass $M_{\rm S2}\simeq 15$
M$_{\odot}$, appears to be a main sequence star, orbiting the
Galactic center black hole with a Keplerian period of $\simeq 15$
yrs. This has allowed \cite{ghezetal} to estimate for the massive
black hole in SgrA$^*$ the mass of $M_{\rm SgrA}\simeq 4.07\times
10^6 ~M_{\odot}$. The orbital parameters of the S2-SgrA$^*$ binary
system are listed in Table \ref{tab1}.

\begin{table}[h]
\begin{center}
\begin{tabular}{ll}
\hline
\multicolumn{2}{c}{{\rm ~~~S2-SgrA$^*$ orbital parameters~~~~}}\vspace{0.1 cm}\\
\hline $M_{\rm SgrA}$&\hspace{1.5 cm}$4.07 \times 10^ 6~M_{\odot}$
\vspace{0.1 cm}\\
$M_{\rm S2}$ & \hspace{1.5 cm}$15~M_{\odot}$ \vspace{0.1 cm}\\
$R_{\rm S2}$ & \hspace{1.5 cm}$5.8~R_{\odot}$ \vspace{0.1 cm}\\
$d$& \hspace{1.5 cm}$4.87 \times 10^{-3}$~pc\vspace{0.1 cm}\\
$e$& \hspace{1.5 cm}$0.87$\vspace{0.1 cm}\\
$P$&\hspace{1.5 cm}$15.78$~yr\vspace{0.1 cm}\\
$i$& \hspace{1.5 cm}$47.3$~deg\\
\hline
\end{tabular}
\end{center}
\caption{The masses $M_{\rm SgrA}$ and $M_{\rm S2}$ of the galactic
center black hole and S2 orbiting star are given. The remaining
orbital parameters are the S2 star radius $R_{\rm S2}$, the
semi-major axis $d$, the eccentricity $e$, the orbital period $P$
and the inclination angle $i$. Data are taken from Ghez et al.
\cite{ghezetal}.} \label{tab1}
\end{table}
The relatively short orbital period of the S2 star allows in
principle to easily attempt an observational campaign to look for
genuine relativistic effects just like the orbital periastron
shift. Since the amount of periastron advance strongly depends on
the compactness of the central body, the detection of such an
effect will give information about the nature of the dark object
hosted in the center of the Galaxy. Hence, it becomes interesting
to estimate the periastron shift for the S2-SgrA$^*$ binary system
by assuming that the central gravitational field source is one of
the object described in the previous sections. The results
(expressed in arcseconds per revolution) are shown in Table
\ref{tab2} and \ref{tab3} for fixed values of the parameters
characterizing the solutions. 

\begin{figure}
\begin{center}
\includegraphics[scale=0.6]{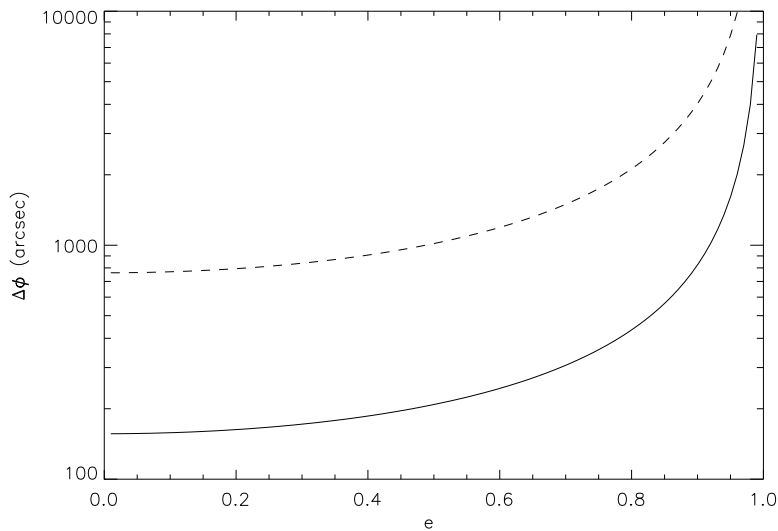}
\end{center}
\caption{The expected periastron shift for an S2-like star as a
function of the orbit eccentricity is shown assuming a central
Schwarzschild black hole with mass $4.07 \times 10^6$ M$_{\odot}$.
Dashed and solid lines are obtained for semi-major axes $d =
10$ mpc and $d =5$ mpc, respectively.} 
\label{fig:1}
\end{figure}

\begin{figure}
\begin{center}
\includegraphics[scale=0.6]{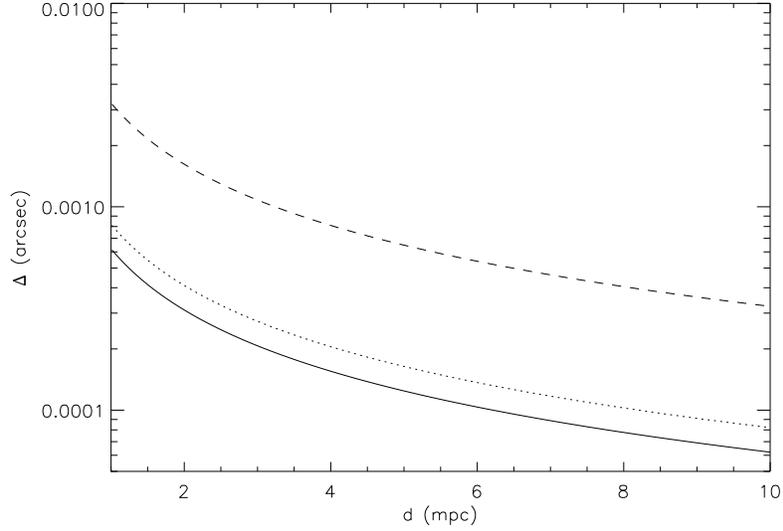}
\end{center}
\caption{The difference $\Delta$ between the expected single
Schwarzschild black hole and the single Curzon particle periastron shifts
(which exactly coincides with that estimated for the corresponding two-body solutions)
is shown as a function of the semi-major axis
$d$. Here, solid, dotted and dashed lines are obtained assuming
$e = 0.1$, $0.5$ and $0.9$, respectively.}
\label{fig:2}
\end{figure}

\begin{figure}
\begin{center}
\includegraphics[scale=0.6]{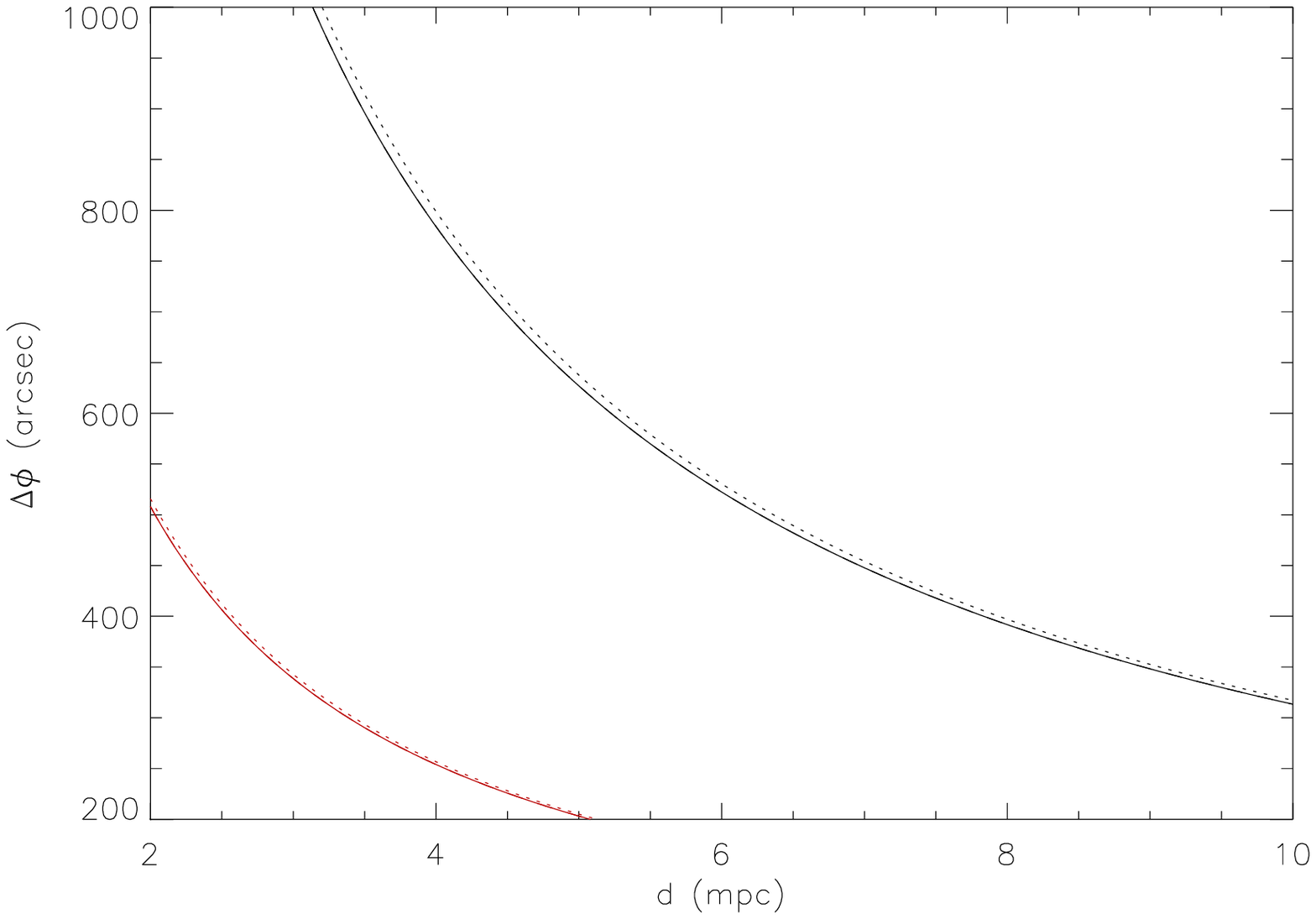}
\end{center}
\caption{The expected periastron shift for an S2-like star as a
function of the orbital semi-major axis is shown. Here, we have
fixed the orbit eccentricity to $e=0.5$ (lower curves) and $e=0.87$
(upper curves). Solid and dashed lines correspond to the
Schwarzschild and maximally Kerr ($a=M_{\rm K}$) black holes cases.}
\label{fig:3}
\end{figure}

Obviously, the previous treatment
also holds for different stars orbiting the galactic center region
and for which the relativistic periastron advance effect might be
clearer. In Figure \ref{fig:1}, the expected periastron shift for
an S2-like star as a function of the orbit eccentricity is shown
assuming a central Schwarzschild black hole with mass $4.07 \times
10^6$ M$_{\odot}$. Dashed and solid lines are obtained for
semi-major axes $d = 10~,5$ mpc, respectively. The difference 
$\Delta$ between the expected single Schwarzschild black hole and
the single Curzon particle periastron shifts (which is equal to that 
estimated for the corresponding two-body solutions, as from equations
(\ref{1ccmen1schw}) and (\ref{2ccmen2schw})) is shown in
Figure \ref{fig:2}. Here, solid, dotted and dashed lines are
obtained assuming  $e = 0.1$, $0.5$ and $0.9$, respectively. In
Figure \ref{fig:3}, instead, we show the expected periastron shift as a
function of the orbital semi-major axis $d$, for fixed values of
the orbit eccentricity $e=0.5$ and $e=0.87$. Solid and dashed lines correspond to the Schwarzschild and
maximally Kerr ($a=M_{\rm K}$) black holes cases, which for the chosen
parameters represent the more favorable ones.


\begin{table}
\begin{center}
\begin{tabular}{|c|c|c|c|c|}
\hline
\rule{0pt}{3ex}$\Delta\phi$ (arcseconds/revolution) & $1\, {\rm CC}$& $1\, {\rm S}$  & $2\, {\rm CC}$ & $2\, {\rm S}$\\
\hline
\rule{0pt}{2ex}${\rm I\,\, order\,\, term}$ & $6.5944\times 10^2$ & $6.5944\times 10^2$& $6.5944\times 10^2$ & $6.5944\times 10^2$ \\
\rule{0pt}{2ex}${\rm II\,\, order\,\, term}$ & $3.4716\times 10^{-1}$ & $3.2822\times 10^{-1}$ & $2.2092\times 10^{-1}$ & $2.1619\times 10^{-1}$ \\
\rule{0pt}{2ex}${\rm I\, +\, II\,\, order\,\, terms}$ & $6.5986\times 10^2$ & $6.5977\times 10^2$ & $6.5966\times 10^2$ & $6.5965\times 10^2$ \\
\hline
\end{tabular}
\end{center}
\caption{The periastron shifts (in arcseconds per revolution)
corresponding to the single Chazy-Curzon particle, single
Schwarzschild black hole, two Chazy-Curzon particles and two
Schwarzschild black holes, respectively, are estimated in the case
of the S2-SgrA$^*$ binary system, whose orbital parameters are
listed in Table \ref{tab1}. Here, we have assumed the S2 star
orbital period to be $15.78$ yrs. In the last two cases, the mass
of each body has been taken equal to half of the total mass
$M=M_{\rm SgrA}$ of the source, and the distance between them has
been fixed equal to $b=3M$. Note also that the second order term
in these two cases can be made vanishing in correspondence of the
choices ${\tilde b}/M\approx9.96$ for 2CC and ${\tilde
b}/M\approx9.89$ for 2S.} \label{tab2}
\end{table}


\begin{table}
\begin{center}
\begin{tabular}{|c|c|c|}
\hline
\rule{0pt}{3ex}$\Delta\phi$ (arcseconds/revolution) & $1\, {\rm K}$ & $1\, {\rm CCrot}$ \\
\hline
\rule{0pt}{2ex}${\rm I\,\, order\,\, term}$ & $6.5944\times 10^2$ & $6.5944\times 10^2$ \\
\rule{0pt}{2ex}${\rm II\,\, order\,\, term}$& $2.3301\times 10^1$ & $-2.2570\times 10^1$ \\
\rule{0pt}{2ex}${\rm I\, +\, II\,\, order\,\, terms}$ & $6.8274\times 10^2$ & $6.3687\times 10^2$ \\
\hline
\end{tabular}
\end{center}
\caption{The periastron shifts (in arcseconds per revolution)
corresponding to the single Kerr black hole as well as to the
single rotating Chazy-Curzon particle are estimated in the case of
the S2-SgrA$^*$ binary system, with the choice of the parameter
$a/M_{\rm K}=1$ and $\alpha=1/6$. In correspondence of this
particular choice of the rotational parameters, the Kerr black
hole and the rotating Chazy-Curzon particle are endowed with the
same total angular momentum $J_{\rm K}=M^2=J_{\rm C_{rot}}$.}
\label{tab3}
\end{table}


\section{Conclusions}

In this paper we have investigated the periastron position advance for
various solutions belonging to the Weyl class of vacuum axially
symmetric solutions to the Einstein field equations, namely those
corresponding to a single Chazy-Curzon particle or Schwarzschild
black hole as well as a pair of them (static), and to a single rotating
Chazy-Curzon particle or Kerr black hole (stationary).
Although differences in the periastron shift among the considered solutions
appear only at second order in the total mass $M$ of the central object,
we have shown that the contribution of the second order term is appreciable
with respect to the first order one, and can discriminate between the
different configurations we have analyzed.

Our results are then applied to the physical system of
the S2-SgrA$^*$ binary system at the galactic center. This research is particularly
timely since within the next years it would be possible to
definitely measure the periastron shift of the S2 star and
therefore to univoquely determine, in principle, the nature of the dark object (usually assumed to be a black hole)
hosted in the center of the Galaxy. However, a caution is needed
since we expect that the galactic center dark object is surrounded
by a compact stellar cluster which may gravitationally interact
with the orbiting star producing an additional periastron shift
which is retrograde with respect to the pure periastron shift
predicted by the General Theory of Relativity. The effects of the
presence of a stellar cluster at the galactic center and the
constraints that present observations can put on its parameters
(i.e. total mass, core radius and central matter density) will be
discussed elsewhere.


\begin{thebibliography}{00}

\bibitem{chazy}
Chazy M., {\it Bull.\ Soc.\ Math.\ France\/}, {\bf 52}, 17 (1924).

\bibitem{curzon}
Curzon H., {\it Proc.\ London\ Math.\ Soc.\/}, {\bf 23}, 477 (1924).

\bibitem{sem1}
Semer\'ak O., Zellerin T. and ${\check {\rm Z}}$\'a${\check {\rm c}}$ek  M.,
{\it MNRAS} {\bf 308}, 691 (1999).

\bibitem{sem2}
Semer\'ak O., ${\check {\rm Z}}$\'a${\check {\rm c}}$ek M. and Zellerin T.,
{\it MNRAS} {\bf 308}, 705 (1999).

\bibitem{weyl}
Weyl H., \textit{Ann.\ Phys.,\ Lpz.} {\bf 54}, 117 (1918).

\bibitem{letelier}
Letelier P.S. and Oliveira S.R., {\it Class.\ Quantum\ Grav.\/},
{\bf 15}, 421 (1998).

\bibitem{scott}
Scott S.M. and Szekeres P.,
{\it Gen. Relativ. Grav.}, {\bf 18}, 557 (1986);
{\it Gen. Relativ. Grav.}, {\bf 18}, 571 (1986).

\bibitem{sokolov}
Sokolov D.D. and Starobinskii A.A., {\it Sov.\ Phys.\ Dokl.\/},
{\bf 22}, 312 (1977).

\bibitem{israel}
Israel W., {\it Phys.\ Rev.\/}, {\bf D15}, 935 (1977).

\bibitem{biniger}
Bini D., Geralico A.,
{\it Int. J. Mod. Phys. D}, {\bf 13}, 983 (2004).

\bibitem{exactsols}
Stephani H., Kramer D., McCallum M.A.H., Hoenselaers C. and Hertl
E., \textit{Exact solutions of Einstein's field equations},
Cambridge University Press, Cambridge (1979).

\bibitem{manko}
Hern\'andez-Pastora J.L., Manko V.S. and Mart\'in J., {\it J.\ Math.\ Phys.\/}, {\bf 34}, 4760 (1993).

\bibitem{ghezetal}
Ghez A.M., Duch$\hat e$ne G., Matthews K. et al., {\it ApJ\/}, {\bf 586}, L127 (2003).


\end{thebibliography}
\end{document}